\documentclass[prb, twocolumn]{revtex4}

\newcommand{\tv}[1]{\t{\v #1}}

\begin{document}

\title{
Spin-triplet $p$-wave pairing in a 3-orbital model for 
iron pnictide 
superconductors
}

\author{Patrick A. Lee}
\author{Xiao-Gang Wen}
\affiliation{
Department of Physics,
Massachusetts Institute of Technology,
Cambridge, MA 02139, USA
}

 \date{\today}

\begin{abstract}
We examine the possibility that the superconductivity in the newly
discovered FeAs materials may be caused by the Coulomb interaction
between $d$-electrons of the iron atoms. We find that when the
Hund's rule ferromagnetic interaction is strong enough, the leading
pairing instability is in spin-triplet $p$-wave channel in the weak
coupling limit.  The resulting superconducting gap has nodal points 
on the 2D Fermi surfaces.  The $\v k$ dependent hybridization of
several orbitals around a Fermi pocket is the key for the appearance
of the spin-triplet $p$-wave pairing.
\end{abstract}

\maketitle


\section{Introduction} 
Recently, a new class of superconductors --
iron based superconductors -- was
discovered.\cite{KHH0612,KWH0896,MZF0828,WMF0821,DZX0826,CLW0890,CWW0803,RYL0883,CHL0895,MCS0896}
The superconducting transition temperature can be as high as
$52$K.\cite{RYL0883} The undoped samples (for example LaOFeAs)
appear to have a spin ordered phase below
$150$K.\cite{DZX0826,CHL0895,MCS0896} The electron
doped\cite{KWH0896} LaO$_{1-x}$F$_x$FeAs and hole
doped\cite{WMF0821} LaO$_{1-x}$Sr$_x$FeAs samples are
superconducting with $T_c\sim 25$K.  
The magnetic field dependence of the specific heat 
in the electron doped material 
suggests the presence of
gapless nodal lines on the Fermi surface.\cite{MZF0828}

It appears that the electron-phonon interaction is not strong enough
to give rise to such high transition temperature.\cite{BDG0803} In
this paper, we will examine the possibility that Coulomb interaction
between $d$-electrons on Fe drives the superconductivity in the
electron/hole doped samples and the spin-order in the undoped
samples.  In this case, 
we find that a $p$-wave spin-triplet superconducting order
with gapless nodal lines is the most likely superconducting order in
the weak coupling limit.  Naively short range repulsion does not
have the requisite $\v k$ dependence to drive $p$-wave pairing. It
turn out that the $\v k$ dependent hybridization of several orbitals
around a Fermi pocket makes the $p$-wave spin-triplet pairing
possible.  The same model is also shown to have spin-density wave
(SDW) order for undoped samples.

\begin{figure}[tb]
\centerline{
\includegraphics[scale=0.37]{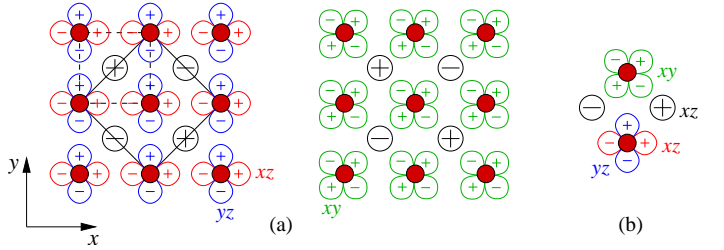}
} \caption{ (Color online) (a) The Fe-As plane and the $d$-orbitals of
Fe.  The filled dots are Fe atoms and the empty dots are As atoms.
The plus and minus signs in the empty dots indicate if the As atom
is above or below the Fe plane.  The large square is the 2D unit
cell.  The dashed square is the reduced unit cell which contains
only one Fe.
The $d_{xz}$, $d_{yz}$ and $d_{xy}$ orbitals of the Fe atoms are described by
the red, blue and green curves respectively.
(b) The $y$ hopping between the $d_{xz}$ or $d_{yz}$
with $d_{xy}$ orbitals.
}
\label{ucellO}
\end{figure}

\begin{figure}[tb]
\centerline{
\includegraphics[scale=0.5]{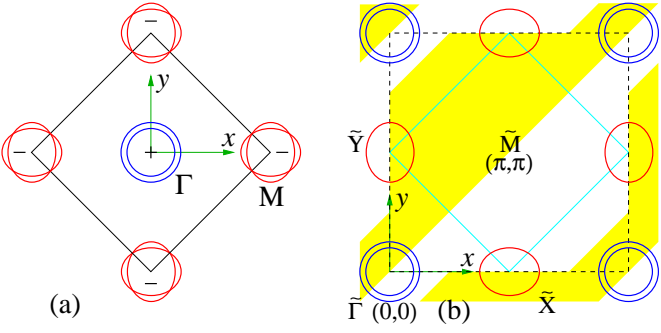}
} \caption{ (Color online) (a) The Fermi surfaces of the Fe
$d$-bands. The plus sign marks the hole pockets and the minus sign
marks the electron pockets. The square is the Brillouin zone. (b)
The extended Brillouin zone (dashed square) for the reduce unit cell
and the positions of the Fermi pockets in the extended Brillouin
zone.  The $\t\Ga$ point has $(\t k_x,\t k_y)=(0,0)$. 
The folding of the extended Brillouin
zone in (b) produces (a). After folding, the $(\t\Ga, \t M)$ and
$(\t X,\t Y)$ in the extended Brillouin zone map into the $\Ga$ and
$M$ in the original Brillouin zone, respectively.
The yellow shading marks the region where the $p$-wave pairing order
parameter may have the same sign. 
}
\label{fs}
\end{figure}

\section{A three-orbital tight-binding model} 

First, let us examine the Fermi surfaces of the iron-based
superconductor.  For concreteness, we will consider the LaOFeAs
sample.  We will assume that the properties of the sample are mainly
determined by the Fe-As planes.  The Fe atoms in a Fe-As plane form a
2D square lattice (see Fig.  \ref{ucellO}).  Due to the buckling of
the As atoms, the real unit cell contains two Fe atoms. The real unit
cell is also a square (see Fig.  \ref{ucellO}).  According to
band structure calculattions, the  Fermi surfaces of the Fe $d$-bands
are formed by two hole pockets at the $\Ga$ point and two electron
pocket at the $M$ point (see Fig.
\ref{fs}a).\cite{SD0829,MSJ0840,KOA0825,HSK0879} All those pockets,
mainly formed by the $d_{xz}$, $d_{yz}$ and $d_{xy}$ orbitals of
Fe,\cite{BDG0803} have similar size, shape, and Fermi velocity.

To understand the mixing of the orbitals near those Fermi pockets, 
we follow Ref. \onlinecite{MSJ0840,KOA0825}
to unfold the band structure to the extended Brillouin zone (BZ).
Ordinarily, such an unfolding of the band structure extends the BZ
superficially and there are certain ambiguities in assigning the
location of each band.  We emphasize that this is not the case here.
The band structure of the extended BZ is uniquely defined due to an
additional symmetry, {\em i.e.} the Fe-As plane is invariant under
$P_zT_x$ and $P_zT_y$, where $T_x$ ($T_y$) is the translation in the
$\v x$ ($\v y$) direction by the Fe-Fe distance and $P_z$ is the
reflection $z\to -z$ (see Fig. \ref{ucellO}).  Thus, if we combine the
translation and the reflection $P_z$, then the electron hopping
Hamiltonian $H$ has a symmetry described by a reduced unit cell with only
one Fe per unit cell (see Fig.  \ref{ucellO}).  
Since $[P_zT_x, H]=[P_zT_y,H]=[P_zT_x,P_zT_y]=0$,
we can use the eigenvalues of $P_zT_x$ and $P_zT_y$ to label
the single-body energy eigenstates:
\begin{equation}
 P_zT_x |\tv k\> = \e^{\imth \t k_x} |\tv k\>,\ \ \ \ \ 
 P_zT_y |\tv k\> = \e^{\imth \t k_y} |\tv k\> ,
\end{equation}
where $\t {\v k}=(\t k_x,\t k_y)$ plays a role of crystal momentum.
We will use such a pseudo crystal momentum to label states in an
energy band. It is the pseudo crystal momenta $\t{\v k}$ that form the
extended Brillouin zone (see Fig. \ref{fs}b) which corresponds to the
reduced unit cell with only one Fe.

Let us work out an explicit example by considering a tight-binding
model involving the $d_{xz}$, $d_{yz}$
and $d_{xy}$ orbitals. First, consider the hopping terms that do not
mix the orbitals:
\begin{align}
H_1 &=-
\sum_{\<\v i\v j\>} [
t_{\v i\v j}^{xz} (c^{xz}_{\v i})^\dag c^{xz}_{\v j}
+t_{\v i\v j}^{yz} (c^{yz}_{\v i})^\dag c^{yz}_{\v j}
\nonumber\\ &\ \ \ \ \ \ \ \ \ \ \ \ \ \ \
+t_{\v i\v j}^{xy} (c^{xy}_{\v i})^\dag c^{xy}_{\v j}
+h.c.]
\end{align}
where $\v i$, $\v j$ label the positions of the Fe.  Since the
$d_{xz}$, $d_{yz}$ and $d_{xy}$ orbitals are eigenstates of $P_z$
reflection, the $P_zT_x$ and $P_zT_y$ symmetries require that  $t_{\v
i\v j}^{xz}$, $t_{\v i\v j}^{yz}$, and $t_{\v i\v j}^{xy}$ only depend
on $\v i-\v j$. 
The mixing term between the $d_{xz}$ and $d_{yz}$ orbitals is given by
\begin{align}
 H_2=-
\sum_{\<\v i\v j\>} [t_{\v i\v j}^{xz,yz} (c^{xz}_{\v i})^\dag c^{yz}_{\v j}
+h.c.],
\end{align}
where $t_{\v i\v j}^{xz,yz}$ also depends only on $\v i-\v j$ as required by
the $P_zT_x$ and $P_zT_y$ symmetries.  On the other hand, the mixing term
between the $d_{xz}$ and $d_{xy}$ orbitals as well as that between the $d_{yz}$ and
$d_{xy}$ orbitals are given by
\begin{align}
 H_3=
\sum_{\<\v i\v j\>} (-)^{i_x+i_y}[
t_{\v i\v j}^{xz,xy} (c^{xz}_{\v i})^\dag c^{xy}_{\v j}
+t_{\v i\v j}^{yz,xy} (c^{yz}_{\v i})^\dag c^{xy}_{\v j}
+h.c.],
\end{align}
where $t_{\v i\v j}^{xz,xy}$, and $t_{\v i\v j}^{yz,xy}$ only depend
on $\v i-\v j$.  We note that the $d_{xz}$ and $d_{xy}$ orbitals have
opposite eigenvalues $\pm 1$ under $P_z$. The $P_zT_x$ and $P_zT_y$
symmetries require the presence of the factor $(-)^{i_x+i_y}$.  Thus,
$c^{xz}$ and $c^{yz}$ with conventional crystal momentum $\v k+\v Q$
can mix with $c^{xy}_{\v k}$ with crystal momentum $\v k$ where $\v
Q=(\pi,\pi)$.

However, in the pseudo crystal momentum $\t {\v k}$ space, only
operators with the same pseudo crystal momentum can mix.  Let 
\begin{align}
\t \Psi_{\t{\v k}}
&= (\t c^{xz}_{\t {\v k}},\t c^{yz}_{\t{\v k}},\t c^{xy}_{\t{\v k}})^T 
\end{align}
be the operators with pseudo crystal momentum $\t {\v k}$,
where
\begin{align}
\t c^{xz}_{\t {\v k}} &\sim 
\sum_{\v i} \e^{-\imth (\tv k+\v Q)\cdot \v i} c^{xz}_{\v i}
\nonumber\\
\t c^{yz}_{\t {\v k}} &\sim 
\sum_{\v i} \e^{-\imth (\tv k+\v Q)\cdot \v i} c^{yz}_{\v i}
\nonumber\\
\t c^{xy}_{\t {\v k}} &\sim 
\sum_{\v i} \e^{-\imth \tv k\cdot \v i} c^{xy}_{\v i}
\end{align}
We see that
the pseudo crystal momentum $\tv k$ and the conventional crystal
momentum $\v k$ are related by $\v k=\tv k$ for the $d_{xy}$ orbital
and $\v k+\v Q=\tv k$ for the $d_{xz}$ and $d_{yz}$ orbitals.
The total hopping Hamiltonian $H=H_1+H_2+H_3$ can be written as
$H=\sum_{\tv k} \t \Psi^\dag_{\tv k} M_{\tv k} \t \Psi_{\tv k}$ with
\begin{align}
 M_{\tv k}&=
\bpm
\eps_{xz}(\tv k) & \eps_{xz,yz}(\tv k) & \eps_{xz,xy}(\tv k) \\
\eps_{xz,yz}(\tv k) & \eps_{yz}(\tv k) & \eps_{yz,xy}(\tv k) \\
\eps^*_{xz,xy}(\tv k) & \eps^*_{yz,xy}(\tv k) & \eps_{xy}(\tv k) \\
\epm .
\end{align}
It is worth noting that the above Hamiltonian and the resulting energy
bands are defined on the extended Brillouin zone (labeled by $\tv k$)
of the reduced unit cell (see Fig. \ref{fs}b).

\begin{figure}[tb]
\centerline{
\includegraphics[scale=0.6]{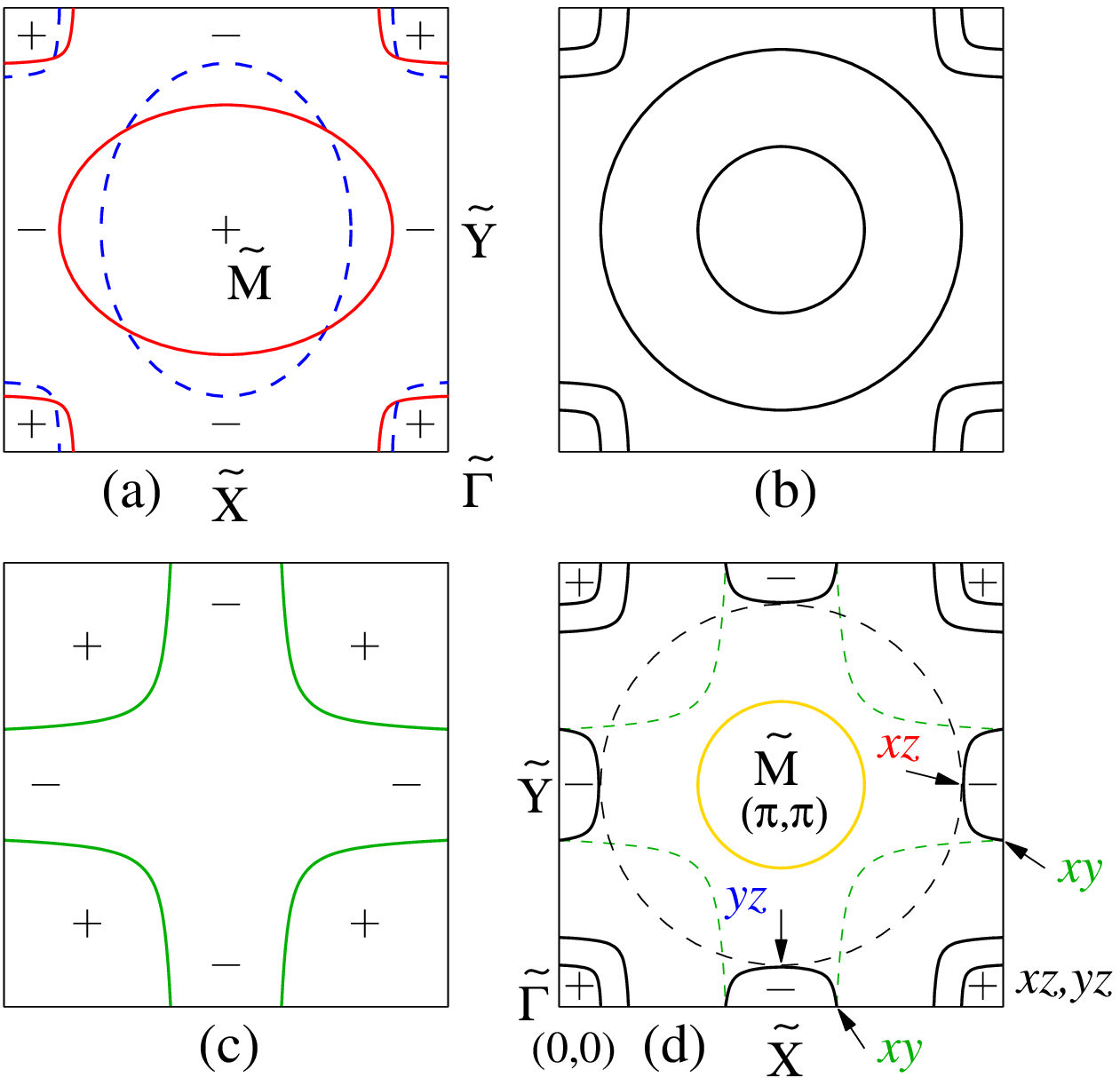}
}
\caption{
(Color online) (a) The zero energy contour of $\eps_{xz}(\tv k)$ (red) and
$\eps_{yz}(\tv k)$ (dashed blue).  The $\pm$ are signs of $\eps_{xz}(\tv k)$ and
$\eps_{yz}(\tv k)$ in the region. The $\t\Ga$ point has $\tv k=0$ 
(b) The hybridization of the $d_{xz}$ and
$d_{yz}$ bands.  The curves are the zero energy contours of the hybridized
bands.  (c) The zero energy contour of $\eps_{xy}(\tv k)$.  (d) The solid
curves are the Fermi surfaces of the three-band tight-binding model as a result
of the hybridization of (b) and (c).
}
\label{bzO}
\end{figure}

The nearest neighbor admixture between $d_{xz}$ and $d_{yz}$ vanishes
by symmetry. Keeping the next nearest neighbor term only, we have $
\eps_{xz,yz}=-2t'_{xz,yz}[\cos(\t k_x+\t k_y)-\cos(\t k_x-\t k_y)]$.
Note that this vanishes at $\t \Ga$, $\t M$, $\t X$ and $\t Y$ points
(see Fig. \ref{bzO}), but is maximal midway between $\t X$ and $\t Y$.
Next, consider the nearest neighbor admixture between $d_{xz}$ or
$d_{yz}$ with $d_{xy}$ along the $y$ direction (see Fig.
\ref{ucellO}b). These orbitals can admix only because of the asymmetry
introduced by the As ions, which has $xz$ symmetry. This implies that
the $d_{xz}$-$d_{xy}$ overlap integral is odd under $x\to -x$ and
vanishes. Only the $d_{yz}$-$d_{xy}$ matrix element survives. 
Also note that under the $180^\circ$ rotation in the $x$-$y$ plane about
the $\v i$ site, $(c^{yz}_{\v i+\v y})^\dag c^{xy}_{\v i} \to
-(c^{yz}_{\v i-\v y})^\dag c^{xy}_{\v i} $.  Thus only the combination
$(c^{yz}_{\v i+\v y})^\dag c^{xy}_{\v i} -(c^{yz}_{\v i-\v y})^\dag
c^{xy}_{\v i}$ that preserves such a symmetry can appear in the
hopping Hamiltonian  which has a form
\begin{align}
 &\ \ \
 \sum_{\v i} \Big[(-)^{i_x+i_y} t_{yz,xy} 
[
(c^{yz}_{\v i+\v y})^\dag c^{xy}_{\v i} 
-(c^{yz}_{\v i-\v y})^\dag c^{xy}_{\v i} 
]
+ h.c. \Big]
\nonumber\\
&= \sum_{\tv k} \Big[ -t_{yz,xy} 
(
\e^{\imth \t k_y}
-\e^{-\imth \t k_y}
)
(\t c^{yz}_{\tv k})^\dag \t c^{xy}_{\tv k} + h.c.
\Big]
\end{align}
This allows us to conclude
that the nearest neighbor $d_{xz}$-$d_{xy}$ and $d_{yz}$-$d_{xy}$
mixing give rise to $\eps_{xz,xy}(\tv k)=-2\imth t_{xz,xy} \sin(\t
k_x)$ and $\eps_{yz,xy}(\tv k)=-2\imth t_{yz,xy} \sin(\t k_y) $.

Now we can see how the 3-orbital model can reproduce the four-pocket Fermi
surface.  Let us begin by assuming that $\eps_{xz}(\tv k)$ lies just below
the Fermi energy (-0.2eV) at $\tv k=\t Y$ and disperses rapidly upward to wards
$\t M$, reaching 2.5eV. From $\t\Ga$, it descends towards $\t X$ where its
energy is $-1.4$eV.  The dispersion is relatively flat along $\t
X$-$\t\Ga$-$\t Y$ and a shallow local maximum appear at $\t\Ga$.  The Fermi
surface correspond to this band is shown in Fig. \ref{bzO}a.  The $d_{yz}$
band is similar except rotated by $90^\circ$.  The reason for the choice of
locating the $d_{xz}+d_{xy}$ band at $\t Y$ (as opposed to $\t X$) will be
explained later.


Next, we turn on the hybridization between $d_{xz}$ and $d_{yz}$. The
hybridization is maximal at $X$ (midway between $\t X$ and $\t Y$) and creates
two sheets which touches at $\t\Ga$ and two Fermi surfaces.  We also find two
concentric small hole pockets at $\t\Ga$ which become the well known hole
pockets at $\Ga$ after folding.

\begin{figure}[tb]
\centerline{
\includegraphics[scale=0.5]{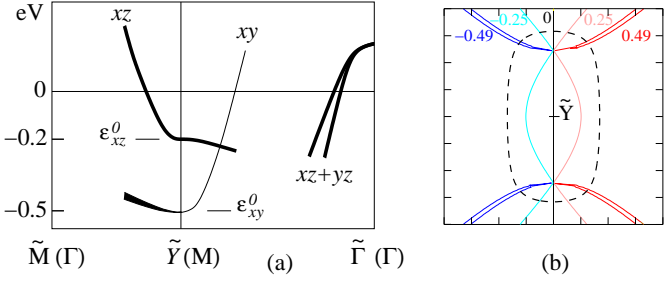}
} \caption{ 
(a) The energy bands near $\t Y$ and $\t \Ga$.
The bands  near $\t Y$ come from the $d_{xz}$ and $d_{xy}$ orbitals.
The thickness of the curve indicates the weight in the $d_{xz}$
orbital. The bands  near $\t \Ga$ came from the $d_{xz}$ and $d_{yz}$
orbitals. 
(b) The contour plot of $\imth u_{\v k}v_{\v k}$
near $\t Y$. The dashed loop is the Fermi surface.
} \label{mix}
\end{figure}

The $d_{xy}$ band $\eps_{xy}(\tv k)$ is assumed to be at
$-0.5$eV at $\tv k=\t X$ and $\tv k=\t Y$, and disperses rapid upwards
toward $\t\Ga$ but rather flat towards $\t M$.  The Fermi surface is
sketched in Fig.  \ref{bzO}c. Now we turn on $\eps_{xz,xy}$ and
$\eps_{yz,xy}$. This gives the electron pockets centered at $\t X$
and $\t Y$ which become the two electron pockets at $M$ after the
folding.  Note that because $\eps_{xz,yz}=0$ at $\t X$ and $\t Y$,
the band at $\t Y$ (or $\t X$) is purely $d_{xz}+d_{xy}$ (or
$d_{yz}+d_{xy}$). Furthermore, near $\t Y$, $\eps_{xz,xy}\sim \sin
\t k_x$.  Thus along $\t Y$-$\t\Ga$, the hybridization is zero and the
$d_{xz}$ and $d_{xy}$ bands cross as shown in Fig.  \ref{mix}a.
Along $\t Y$-$\t M$, the upper band is mostly $d_{xz}$ while the
$d_{xz}$ amplitude in the lower band increase linearly with the
distance from $\t Y$. These features are in agreement with band
calculations.\cite{BDG0803,KOA0825} and are strong evidences that
our assignment of the $d_{xz}$ band is correct.

The final result is an ellipse shaped pocket, with the long axis of
the ellipse pointing in the $y$ direction at $\t Y$ (towards the hole pockets
at $\t\Ga$). The Fermi surface crossing along the long direction is purely
$d_{xy}$ and the short direction is mainly $d_{xz}$ (see Fig. \ref{bzO}d).

The pocket formed between the dashed lines in Fig. \ref{bzO}d are
eliminated for sufficiently strong hybridization. However, within
the 3-orbital model, it is impossible remove the Fermi surface
surrounding $\t M$ (the gold solid loop in Fig. \ref{bzO}d), because
the hybridization elements are all zero at $\t M$. We need a 4th band
which crosses and hybridize with the $d_{xz}+d_{yz}$ band to
eliminate the unwanted Fermi surface. Apart from this, the 4th band
plays no role as far as the remaining two electron and two hole
pockets are concerned.  It is in this sense that we maintain that
the 3-orbital model gives an adequate description of the low energy
Hamiltonian.

\section{The interaction between $d$-electrons}
Next, let us consider the single ion interaction between the electrons on the Fe
$d$-orbitals which is given by
\begin{align}
 H_I &=\frac12 \sum_{\al,\al'}
\int \dd^2 \v x\dd^2\v x'\;
c^\dag_{\al}(\v x)c_{\al}(\v x)
V(\v x-\v x')
c^\dag_{\al'}(\v x')c_{\al'}(\v x')
\nonumber\\
&=\frac12 \sum_{a_1,a_2,a_3,a_4,\al,\al'}
c^\dag_{\al a_1} c^\dag_{\al' a_2} c_{\al'a_3} c_{\al a_4}\times
\\&\ \ \ \ \
\int \dd^2 \v x\dd^2\v x'\;
\phi_{a_1}(\v x) \phi_{a_2}(\v x') V(\v x-\v x')
\phi_{a_3}(\v x') \phi_{a_4}(\v x)
\nonumber 
\end{align}
where $a_1,\cdots,a_4=xz,yz,xy$ label the orbitals, $\phi_a(\v x)$ is the wave
function of the orbitals, and $c_a$ is the electron operator for the
$a$-orbital.
Due to the symmetry of the orbitals, $a_i$ must appear in pairs
and the above can be rewritten as
\begin{align}
\label{HIUJ}
H_I&= \frac12 U_1\sum_a
c^\dag_{\al a} c^\dag_{\bt a} c_{\bt a}c_{\al a}
+
\frac12 U_2\sum_{a\neq b}
c^\dag_{\al a} c^\dag_{\bt b} c_{\bt b}c_{\al a}
\nonumber\\&
+\frac12 J \sum_{a\neq b}
c^\dag_{\al a} c^\dag_{\bt b} c_{\bt a}c_{\al b}
+\frac12 J\sum_{a\neq b}
c^\dag_{\al a} c^\dag_{\bt a} c_{\bt b}c_{\al b}
\end{align}
where
$U_1=U_{aa}=U_{bb}$, $U_2=U_{ab}=U_1-2J$,\cite{CNR7845} and
\begin{align}
 U_{ab}&=\int \dd^2 \v x\dd^2\v x'\;
 \phi_{a}(\v x) \phi_{b}(\v x') V(\v x-\v x') \phi_{b}(\v x') \phi_{a}(\v x) ,
\nonumber\\
 J&=\int \dd^2 \v x\dd^2\v x'\;
 \phi_{a}(\v x) \phi_{b}(\v x') V(\v x-\v x') \phi_{a}(\v x') \phi_{b}(\v x) .
\end{align}

Here the $U_1$ term represents the Coulomb interaction between electrons on
same orbital.  The $U_2$ term represents the interaction between electrons on
different orbitals.  The first $J$ term is the exchange effect that favor
parallel spins on the same Fe and is responsible for the Hund's rule. The
second $J$ term is pair hopping between different orbitals.

\section{Pairing instability} The above $d$-electron interaction may cause a
pairing instability.  One possible pairing instability is the spin-triplet
pairing between $\t X$ and $\t Y$ electron pockets.\cite{DFZ0882} In this
paper, we will consider a different type of pairing -- the spin-triplet
$p$-wave pairing within the same Fermi pocket.

First, let us consider the $p$-wave pairing on the Fermi pocket near the $\t Y$
point (see Fig. \ref{bzO}).  Such a Fermi pocket is a mixture of the $d_{xz}$
and $d_{xy}$ orbitals.  Let $\psi_{xz}$ and $\psi_{xy}$ be the electron operators in
the continuum limit in the $d_{xz}$ and $d_{xy}$ orbitals near the $\t Y$ point.
The electron operator $\psi$ near the pocket at the $\t Y$ point (see Fig.
\ref{fs}a) is a mixture of $\psi_{xz}$ and $\psi_{xy}$.

The Hamiltonian has the $P_zP_x$, $P_zP_y$, $P_{xy}$, $P_zT_x$, and
$P_zT_y$ symmetries, where $P_x: x\to -x$ and $P_y: y\to -y$ are
reflections about a Fe atom. These symmetry dictates certain form of
the continuum Hamiltonian. Alternatively, we can expand the
tight-binding picture near $\t Y$ to obtain the following form which
respect all the symmetry requirements
\begin{align}
 H_0 &=
(t_1 k_x^2+t_2 k_y^2+\eps^0_{xz}) \psi^\dag_{xz} \psi_{xz}
\\ &\
+(\t t_1 k_x^2+\t t_2 k_y^2+\eps^0_{xy}) \psi^\dag_{xy} \psi_{xy}
+ t_3 k_x (\imth \psi^\dag_{xy} \psi_{xz} +h.c.)
\nonumber 
\end{align}
Here $\v k$ is the pseudo crustal momentum
measured from the $\t Y$ point.  From Fig. \ref{mix}(a), we
see that the $d_{xz}$ band is flat along the $y$-axis, which implies 
$t_1\gg t_2$. Similarly $\t t_2\gg \t t_1$.  In the following, we will ignore
$t_2$ and $\t t_1$.

The electrons have two bands with energies 
\begin{align}
 E_{\pm}(\v k) =
\eps_0\pm\sqrt{\eps_2^2+\eps_3^2} ,
\end{align}
 where 
\begin{align}
\eps_0 &=\frac12 ( t_1 k_x^2+\t t_2 k_y^2+\eps^0_{xz}+ \eps^0_{xy}), 
\nonumber\\
\eps_3 &=\frac12 ( t_1 k_x^2-\t t_2 k_y^2+\eps^0_{xz}- \eps^0_{xy}) ,  
\nonumber\\
\eps_2 &= t_3 k_x  .  
\end{align}
$\psi_{xz}$ and
$\psi_{xy}$ are related to the electron operator $\psi$ in the upper band
$E_+$ as 
\begin{align}
\psi_{xz}(\v k)=u_{\v k} \psi(\v k), \ \ \ \ \
\psi_{xy}(\v k)=v_{\v k} \psi(\v k), 
\end{align}
where
\begin{align}
u_{\v k}&=\frac{\imth \eps_2}{
\sqrt{ 2 \eps_2^2+2 \eps_3^2 - 2 \eps_3 \sqrt{\eps_2^2+\eps_3^2}} } , 
\nonumber\\
v_{\v k}&=\frac{\eps_3 - \sqrt{\eps_2^2+\eps_3^2} }{
\sqrt{ 2 \eps_2^2+2 \eps_3^2 - 2 \eps_3 \sqrt{\eps_2^2+\eps_3^2}} }.
\end{align}

To obtain the pairing interaction near the $\t Y$ point in the spin-triplet
channel we set $\al=\bt=\al'=\bt'=\up$ in \eq{HIUJ}.  We find that the first
and the fourth terms in \eq{HIUJ} vanish.  The second and the third terms
become
\begin{align}
\sum_{\v k_1,\v k_2} V_{\t Y}(\v k_2,\v k_1)
[\psi_\up (\v k_2) \psi_\up (-\v k_2)]^\dag
\psi_\up (\v k_1) \psi_\up (-\v k_1)
\end{align}
in the spin-triplet and $(\v k,-\v k)$ pairing channel, where
the effective pairing interaction of $\psi$ is
\begin{align}
\label{Vk1k2M}
V_{\t Y}(\v k_2,\v k_1)=- (J-U_2)
u^*_{\v k_2} v^*_{-\v k_2}
u_{\v k_1} v_{-\v k_1}
\end{align}
{}From the effective pairing interaction, we can obtain
a dimensionless coupling constant\cite{SLH8794}
\begin{equation}
 \la=-\frac{
\int \frac{\dd \si_k}{(2\pi)^2 |v_{\v k}|}
\int \frac{\dd \si_{k'}}{(2\pi)^2 |v_{\v k'}|}
g^*(\v k) V(\v k,\v k') g(\v k')
}{
\int \frac{\dd \si_{k}}{(2\pi)^2 |v_{\v k}|}
|g(\v k)|^2
}
\end{equation}
where $\int \dd \si_k$ is the integration over the $\t Y$ Fermi surface and
$v_{\v k}$ is the Fermi velocity.
The function $g(\v k)$ is a square harmonics which
describes the shape of the superconducting gap, {\sl e.g.} $g(\v k)=1$
corresponds to an $s$-wave and $g(\v k)=\sin k_x$ or $\sin k_y$ a $p$-wave
superconductor.  The superconducting transition temperature $T_c$ is given by $
T_c = \Om \e^{-1/\la}$, where $\Om$ is of order of the Fermi energy of the
pocket: $\Om \sim 0.2$eV.

{}From \eq{Vk1k2M}, we find that, when $J>U_2$, the pairing interaction $V_{\t
Y}(\v k_2,\v k_1)$ induces a pairing $g(\v k)$ that have the same symmetry as
$u_{\v k}v_{\v k}$.  Because $u_{\v k}v_{\v k}$ is odd in $k_x$ (see Fig.
\ref{mix}b), the induced pairing is in $p$-wave channel $g(\v k)=\sin k_x$.
Thus when $J>U_2$, the $d$-electron interaction will cause a spin-triplet
$p$-wave pairing.  Note that the location of the node depends on the choice of
$\sin k_x$ versus $\sin k_y$, which in turn hinges on our assignment of the
orbital to be $xz$-like near $\t Y$.  Triplet pairing has been proposed
earlier,\cite{XMY0882} and fully gapped states such as $p_x+\imth p_y$ were
suggested. In contrast, after including the $\v k$-dependent orbital-mixing,
we find the on-site ferromagnetic interaction to favor a particular nodal
$p_x$ and $p_y$ states in the $\t Y$ and $\t X$  valleys respectively.  

As long as $\eps_F> \eps^0_{xz}$, we see from Fig. \ref{mix}a that the Fermi
surface changes its character from pure $d_{xy}$ to mostly $d_{xz}$ as
a function of angles.  $|u_{\v k}|^2$ and $|v_{\v k}|^2$  must cross at some
angles where $|u_{\v k} v_{\v k}|$ takes its peak value 1/2 (see Fig.
\ref{mix}b).  Thus the effect of $|u_{\v k} v_{\v k}|$ or $\la$ is relatively
insensitive to doping provided that $\eps_F> \eps^0_{xz}$.  Since the
density of states is also independent of doping in 2D, this explains why $T_c$
is somewhat insensitive to doping and may extended to the hole doped
side,\cite{WMF0821} except near zero doping. There the $U_1$ term will drive
an SDW instability due to the nesting between the electron and the hole
pockets.\cite{ML0886}$^,$\cite{MSJ0840} 

We would like to mention that since $E(\v k)=E(-\v k)$, our intra-pocket
$p$-wave pairing appears even when the attraction $J-U_2$ is weak.  When the
attraction  $J-U_2$ is strong enough to overcome the inter-pocket energy
splitting ($\sim 0.05$eV), our model also has the instability in the
spin-triplet inter-pocket pairing channel proposed in Ref.
\onlinecite{DFZ0882}.  Since the intra-pocket effective attraction is reduced
by the matrix elements as shown in \eq{Vk1k2M}, the inter-pocket pairing may be
stronger in large $J-U_2$ limit while the intra-pocket pairing is stronger in
small $J-U_2$ limit.

Similarly, we can consider the spin-triplet pairing on the Fermi
pocket near the $\t\Ga$ point (see Fig. \ref{fs}b).  Such a Fermi
pocket is a mixture of the $d_{xz}$ and $d_{yz}$ orbitals.  The
dispersion and the $d_{xz}$-$d_{yz}$ mixing near $\Ga$ can be
determined from the symmetry consideration.  The crucial difference is
that the hybridization matrix element is now proportional to $k_xk_y$.
After a similar calculation, we find that the spin-triplet pairing
potential $V_{\t\Ga}(\v k_2,\v k_1)$ satisfy $ V_{\t\Ga}(\v k_2,\v
k_1)= V_{\t\Ga}(\v k_2,-\v k_1)= V_{\t\Ga}(-\v k_2,\v k_1) $. A
$p$-wave pairing will result in a vanishing dimensionless coupling
$\la=0$.  Thus the Coulomb interaction in the $d$-orbitals does not
induce spin-triplet $p$-wave pairing on the two pockets near $\Ga$
even when $J>U_2$.  We see that the $d_{xz}$-$d_{xy}$
($d_{yz}$-$d_{xy}$) mixing at $\t Y$ ($\t X$) in the 3-orbital model
is crucial for the appearence of our $p$-wave instability.  A
2-orbital model has the wrong Fermi surface topology in that one hole
pocket is located at $\t \Ga$ and $\t M$ in Fig. \ref{fs}(b), instead
of both being at $\t \Ga$.\cite{RQL0813}  Although the 2-orbital model
may allow certain superconducting states,\cite{HCW0846,L0836} it does
not have the proper symmetry and the orbital mixing to generate the
$p$-wave pairing proposed here.

\section{Conclusion} 
In this paper, we examine the possibility that
Coulomb interaction between the electrons in the Fe $d$-orbitals may
induce a superconducting phase.  We find that when the Hund's rule
ferromagnetic interaction on Fe is strong enough, \ie when $J>U_2$,
the  Coulomb interaction can induce pairing instability.  In the
weak coupling limit, the leading instability is found to be a
spin-triplet $p$-wave pairing on the electron pockets at $M$ (or at
$\t X$ and $\t Y$ in the extended Brillouin zone).  Although the
Coulomb interaction can only directly induce pairing on the electron
pockets at $M$, the proximity effect will lead to a $p$-wave pairing
on the hole pockets at $\Ga$.  So the resulting superconducting
state has node lines on the 3D Fermi surfaces.  

The spin triplet pairing order paraneter is a compex vector $\v d$.  The
relative phases and relative orientations of the two order parameters $\v d_{\t
X}$ and $\v d_{\t Y}$ on the two pockets $\t X$ and $\t Y$ can have 
interesting relations which cannot be determined from the linear response
calculation adopted here. One possible distribution of the phases is given in
Fig.  \ref{fs}b, where $p$-wave pairing gap is positive in the shaded regions
and negative in unshaded regions.


At the atomic level, $U \approx 3$ to 4~eV and $J \approx 0.7$~eV, so $U_2 - J$ is positive.  However, in a tight binding model involving only the Fe $d$-orbitals, the As orbitals have been projected out and the appropriate $U$ and $J$ are those corresponding to the Wannier orbitals which are much more extended than the atomic $d$ orbitals.  A recent estimate by Anisimov {\em et al.}\cite{Anisimov} found the average $U$ to be strongly renormalized down to 0.8~eV while $J$ remains large at 0.5~eV.  These are the more appropriate bare parameters in a 3-band model.  Furthermore, 
in a crystal, the strong hopping leads to extended quasiparticles
near the Fermi surface. The on-site $U_2$ and $J$ will induce effective
interaction $U_2^*$ and $J^*$ between those quasiparticles. The term induced
by $U_2$ will remain short ranged. However, $J$ may induce a long range couple
because parallel spin configuration favors hopping, \ie Hund's rule and
hopping are compatible.  Since the Fermi pockets are small, the effective
interaction is given by $U^*_2(q)-J^*(q)$ with $q\sim k_F \ll 1/a$. A long
range $J$-coupling enhances $J^*(q=0)$ and can potentially lead to a sign
change and a net effective attraction.

Since the original submission of this paper the experimental situation has evolved rapidly.  Here we attempt a brief summary of the relevant experimental data.  The issue of whether gap nodes exist remains open to debate.  Photoemission data on Ba$_{1-x}$K$_x$Fe$_2$As$_2$ indicate an almost isotropic gap on the hole pocket in this hole doped material.\cite{Ding}  It was recently found that the magnetic field dependence of the specific heat in this material is linear,\cite{Mu} in contrast with the $\sqrt{H}$ behavior taken as evidence for nodes in the electron doped material.\cite{MZF0828}  At the same time, the nuclear spin relaxation rate $\frac{1}{T_1}$ fits the $T^3$ law over three decades in La(O$_{1-x}$Fe$_x$)FeAs.\cite{Nakai}  Thus at the moment, existing data seem to point to gap nodes in electron doped materials and their absence in hole doped materials.  We note that, because the size of the Fermi pockets are small, it is usually advantageous to hide the nodes in the $k$ space between Fermi pockets.  In order to produce a gap node on the Fermi surface of the small pocket, one needs an effective pairing potential which varies rapidly on the scale of the small pocket.  Our theory is one of the few that will do this, and we rely on the rapid $k$ dependence of the hybridization matrix element.  The message that wavefunction and matrix elements may play an important role and must therefore be handled properly has validity beyond the special $p$-wave pairing scenario described here.

What about singlet vs. triplet pairing?  The Knight shift is probably the best way to answer this question.  In a triplet superconductor, the spin contribution to the Knight shift drops below $T_c$ to zero if the magnetic field is parallel to the $\v d$ vector, but remains unchanged if it is perpendicular.  If the $\v d$ vector is free to rotate, it will turn perpendicular to $\v H$ and no change in Knight shift is predicted.  This is apparently the case for Sr$_2$RuO$_4$.  On the other hand, if $\v d$ is locked to the lattice, we expect to see in a polycrystalline sample a drop of 2/3 of the value compared with singlet pairing.  Without accurate knowledge of the orbital contribution to the Knight shift, this is hard to distinguish.  Thus NMR on single crystals is needed to settle this question.  As of this writing, the only single crystal data available is from Ning {\em et al.}\cite{Ning} on BaFe$_{1.8}$Co$_{0.2}$As$_2$.  The data does not support triplet pairing in that a drop in the Knight shift is seen for field directions both parallel and perpendicular to the plane.  On the other hand, a recent paper by Nakai {\em et al.}\cite{Nakai2} on the FeP system La$_{0.87}$Ca$_{0.13}$FePO shows that the magnetic behavior is quite different from the FeAs system.  The Knight shift increases with decreasing temperature, indicative of ferromagnetic fluctuations above $T_c$ and $\frac{1}{T_1T}$
shows a very unusual increase below $T_c$.  The authors speculate that
magnetic fluctuations associated with triplet pairing may be
responsible for the increase.  The experimental situation remains in
flux and it may be possible that different pairing scenarios may be
competing and win out in different materials.

This research is supported by DOE grant DE-FG02-03ER46076 (PAL) and by NSF
Grant DMR-0706078 (XGW).


\end{document}